\documentclass[pre,twocolumn,superscriptaddress,nobibnotes]{revtex4-2}
\usepackage{multirow}
\usepackage{rotating}
\usepackage{booktabs,tabularray}
\usepackage{amsmath}
\usepackage{blindtext}
\usepackage{mathrsfs}
\usepackage[protrusion=true,expansion=true]{microtype}
\usepackage{times}
\usepackage{bm,xcolor,braket, tikz}
\usepackage{graphicx}
\usepackage[normalem]{ulem}

\usepackage{hyperref}
\hypersetup{colorlinks,linkcolor={blue},citecolor={blue},urlcolor={blue}}
\usepackage{amssymb,physics}
\usepackage[utf8]{inputenc}
\usepackage[english]{babel}
\usepackage{bm}
\makeatletter
\def\graphicscale{\twocolumn@sw{0.3}{0.4}}
\def\graphicthreescale{\twocolumn@sw{0.3}{0.4}}
\newcommand{\rev}[1]{{#1}}
\definecolor{bblue}{rgb}{0.36, 0.54, 0.66}

\newcommand{\be}{\begin{equation}}
\newcommand{\ee}{\end{equation}}
\newcommand{\bea}{\begin{align}}
\newcommand{\eea}{\end{align}}
\newcommand{\de}{\partial}
\newcommand{\E}{\mathbb{E}}
\usepackage{float}
\usepackage{dsfont}
\usepackage{mathrsfs}
\usepackage{comment}
\usepackage{soul}

\newcommand{\titleinfo}{
Large deviations of density fluctuations in the boundary driven \\ Quantum Symmetric Simple Inclusion Process.
}
\begin{document}
\title{\titleinfo}
\newcommand{\nn}{\nonumber\\[4pt]}
\author{Denis Bernard}
\affiliation{Laboratoire de Physique de l’\'Ecole Normale Superieure, CNRS, ENS \& Universit\'e PSL,
Sorbonne Universit\'e, Universit\'e Paris Cit\'e, 75005 Paris, France}
\author{Tony Jin}
\affiliation{Universit\'e Côte d’Azur, CNRS, Centrale Med, Institut de Physique de Nice, 06200 Nice, France}
\author{Stefano Scopa$^1$}
\email{stefano.scopa@phys.ens.fr}
\author{Shiyi Wei$^2$}

\date{\today}
\begin{abstract}
We consider the boundary driven Quantum Symmetric Simple Inclusion Process (QSSIP) which describes a one-dimensional system of bosonic particles with stochastic nearest-neighbor hopping, modeled as a Brownian motion, with gain/loss processes at the endpoints of the chain driving the system out-of-equilibrium. Although the averaged QSSIP dynamics differs from that of the Quantum Symmetric Simple Exclusion Process (QSSEP)~--~the analogous system where bosons are replaced by fermions~--~we show that, paradoxically, the  dynamics of their matrices of two-point functions, along with all their fluctuations, coincide. 
In contrary, the underlying classical models differs significantly, as the bosonic statistics allow the inclusion of multiple particles at the same site, in contrast to (symmetric) simple exclusion processes (SSEP). We provide an exact derivation of the large deviation function of density fluctuations in QSSIP and, as a consequence, in the classical inclusion process (SSIP) by exploiting its quantum formulation.
Remarkably, our study highlights that, both in QSSEP and QSSIP, fluctuations of the local densities are typically  classical, i.e. the cumulant generating functions of the local densities are asymptotically self-averaging and converge toward those of the classical SSEP and SSIP, \textit{realization-wise}. This provides a test of the conjectured almost sure classical behavior of transport fluctuations, at leading order in the system size, in noisy diffusive quantum many-body systems.

\end{abstract}

\maketitle
\section{Introduction}
Understanding non-equilibrium physics remains a central challenge in statistical physics, with exact solutions providing invaluable insights into the underlying mechanisms. Classical stochastic processes, such as the Symmetric Simple Exclusion Process (SSEP) \cite{Kipnis1989, Spohn1991, Kipnis1999}, have served as paradigmatic models, allowing for the derivation of exact large deviation functions through sophisticated combinatorial and algebraic techniques, see e.g.~\cite{derrida1993exact,Derrida2001,Derrida2007,bodineau2004current,Mallick2015}. These studies have significantly enhanced our understanding of fluctuations \cite{gallavotti1995, jarzynski1997, crooks1999}, rare events \cite{touchette2009large,lazarescu2017} and time-reversal symmetry \cite{maes1999, maes2003time}, leading to the development of a macroscopic fluctuation theory for classical diffusive transport \cite{bertini2002macroscopic, bertini2001fluctuations}. However, extending such results to quantum systems remains a major challenge and continues to be an active area of research.

In the quantum realm, interests have initially been focused on transport phenomena, where significant progress has been made in the case of ballistic transport. Generalized hydrodynamics has emerged as a powerful framework~\cite{Bertini2016,C-Alvaredo2016}, enabling the exact description of the integrable models far from equilibrium. In particular, it has provided precise characterizations of local conserved densities and their correlation functions, marking a substantial advancement in the field, see~\cite{Alba2021,Bouchoule2022} for a review.  
For quantum diffusive systems, our understanding remains considerably more limited, especially concerning coherent phenomena beyond transport \cite{Bernard2021}. Modeling such systems by charge conserving quantum circuits has provided a mean-field hydrodynamics picture of entanglement spreading in diffusive chaotic systems \cite{Gullans2019}. In parallel, the Quantum Symmetric Simple Exclusion Process (QSSEP), a fermionic model with stochastic hoppings and Lindblad drivings at the boundaries, has been introduced to gain information on --~and to model~-- coherent fluctuations in quantum diffusive systems out-of-equilibrium \cite{Bauer2019,Bernard2019}.
This framework not only realizes a quantum extension of the classical SSEP but also reveals subtle patterns in coherent fluctuations and deep connections with free probability theory~\cite{biane2022,Hruza2023}. The averaged dynamics of QSSEP reproduces that of the classical SSEP, enabling a different route towards the exact computations of large deviation function for density fluctuations~\cite{Bauer2024}.
Moreover, various studies have demonstrated that the fluctuations of coherence in diffusive metals \cite{LeeLevitovYuPRBUniversalstatistics,RocheDerridaCurrentfluctuationsinSSEP,HruzaJinAndersonQSSEP} can be quantitatively described using the (Q)SSEP framework. These findings raise an important question: To what extent can the exact results derived for quantum exclusion processes, or random quantum circuits, be generalized to establish a universal framework for understanding quantum diffusion and its fluctuations? Such framework would provide an extension of the macroscopic fluctuations theory~\cite{bertini2002macroscopic} to the quantum realm~\cite{Bernard2021}.
Within this framework, it is has been conjectured \cite{Hruza2023} that, to leading order, fluctuations of transport phenomena in noisy quantum diffusive systems are essentially classical in nature. More precisely, this conjecture claims that, in noisy quantum diffusive systems, the quantum fluctuations of transport phenomena satisfy a large deviation principle, \textit{realization-wise}, and that their large deviation functions are, to leading order in the system size, realization independent and equal to that of the associated classical processes defined by the mean dynamics. Hint for the validity of this conjecture has been recently gained in charge conserving random circuits \cite{FCSMFTDeNardis}. Of course, this conjecture refers only to the fluctuations of the transport phenomena but not to those  of quantum coherences which are beyond classicallity.
Addressing this question is challenging, and a definitive answer remains still far on the horizon.

In this work, we take a step in this direction by exploring a bosonic analog of QSSEP characterized by stochastic hoppings and boundary Lindblad driving terms. {While the quantum model exhibits striking similarities with QSSEP, we show that, on average, its dynamics reduces to the Symmetric Simple \textit{Inclusion} Process (SSIP).
Contrary to SSEP, in which an exclusion principle forbids multiple occupancy of any sites, the SSIP dynamics favors such multiple occupancy. See Fig.~\ref{fig:SIP}. 
We study the quantum and statistical fluctuations of density in QSSIP. We prove that, as usual in the thermodynamic limit, those fluctuations are rare events but that they satisfy a large deviation principle. We furthermore prove that, as indeed postulated \cite{Hruza2023} by the conjecture discussed above, the large deviation principle holds almost surely realization-wise and that the large deviation function is that of the classical averaged process SSIP. This almost sure convergence to the classical large deviation function relies on properties of ensembles of structured random matrices recently introduced in \cite{Bernard2024} as a by product of studies of quantum exclusion processes. Through the correspondence between QSSIP and its classical analog, our computation yields an exact formula for the large deviation function for density fluctuations of the SSIP which, to the best of our knowledge, has not been solved in the literature before. The arguments we provide for the almost sure convergence of the large deviation function in QSSIP applies similarly to QSSEP -- as in both cases it relies on properties of structured random matrices \cite{Bernard2024}. As a consequence, the conjecture about the \textit{typical classical nature} of transport fluctuations also tested in QSSEP. Since the averaged dynamics in QSSEP and QSSIP differ, although the dynamics of the matrix of two-point function coincide, comparing the QSSEP and QSSIP allows a finer test of the conjecture: the large deviation function for transport phenomena is indeed that of the classical process specified by the mean dynamics.

\begin{figure}[t]
    \centering
    \includegraphics[width=\columnwidth]{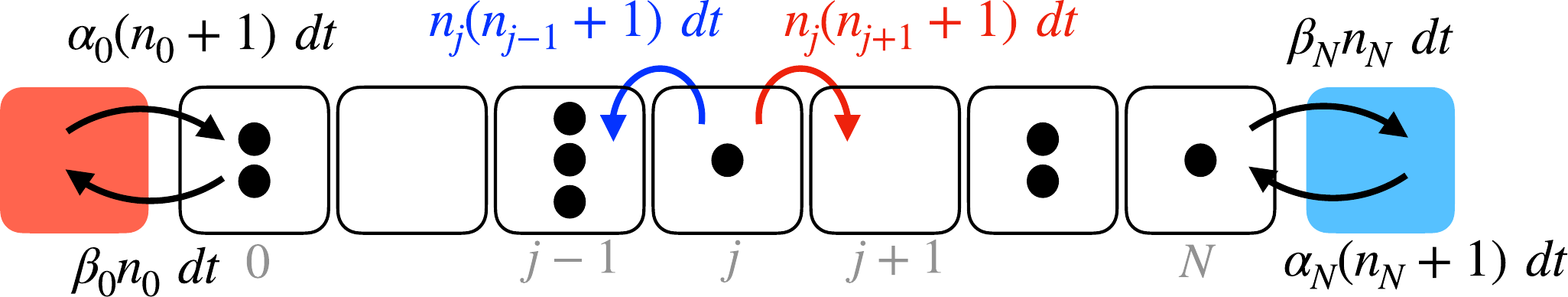}
    \caption{Illustration of the mean classical process in Eqs.~\eqref{eq:cl-bulk} and \eqref{eq:cl-bdry}. A classical configuration of the chain with sites occupations $n_j$ is updated with transition probabilities depending on the occupations of the departure and arrival sites.}
    \label{fig:SIP}
\end{figure}

\section{The model and its steady statistics}

We consider a one-dimensional quantum chain with nearest neighbor stochastic hoppings, whose bulk dynamics is  generated by the stochastic Hamiltonian 
\be\label{eq:sipH}
d\hat{H}_t=\sum_{j=0}^{N-1} \left(\hat{a}^\dagger_{j+1}\hat{a}_{j} dW^j_t + \hat{a}_j^\dagger \hat{a}_{j+1} d\overline{W}^{j}_t\right).
\ee
Here, $\hat{a}^\dagger_j$, $\hat{a}_j$ are standard bosonic creation and annihilation operators acting on site $j\in[0,N]$, satisfying $[\hat{a}_j,\hat{a}^\dagger_k]=\delta_{j,k}$ and zero otherwise.  $W^j_t$, $\overline{W}^j_t$ are pairs of complex conjugated Brownian motions, with non-vanishing quadratic variations $dW^j_t d\overline{W}^k_{s}=\delta_{j,k} \delta_{t,s} dt$. At the endpoints of the chain, we couple our stochastic quantum chain to dissipative couplings. Assuming the interaction between the chain and the two reservoirs to be Markovian, the coupling to the external leads can be modeled by Lindblad terms, resulting in the equation of motion for the system's density matrix
\be\label{eq:lind}
d\hat\rho_t=-i[d\hat{H}_t,\hat\rho_t]-\frac12[d\hat{H}_t,[d\hat{H}_t,\hat\rho_t]]+{\cal L}_\text{bdry}(\hat\rho_t) dt.
\ee
The first two terms on the r.h.s. are obtained from expanding the bulk unitary flow generated by the Hamiltonian \eqref{eq:sipH}, that is $e^{-id\hat{H}_t}\hat\rho_t e^{id\hat{H}_t}$. The boundary term is made of the sum of injection/extraction Lindblad super-operators acting at the endpoints of the chain,
\begin{align}
{\cal L}_\text{bdry}(\bullet)=\sum_{p\in\{0,N\}}\Big[ &\alpha_p \Big(\hat{a}^\dagger_p \bullet \hat{a}_p -\frac12\{\hat{a}_p\hat{a}^\dagger_p,\bullet\}\Big)\nonumber\\
& +\beta_p \Big(\hat{a}_p \bullet \hat{a}^\dagger_p -\frac12\{\hat{a}^\dagger_p\hat{a}_p,\bullet\}\Big)\Big]
\end{align}
with $\alpha_p$ and $\beta_p$ denoting the injection and extraction rates, respectively, at the boundary site $p\in\{0,N\}$ of the chain. Realization-wise, the dynamics induced by Eq.~\eqref{eq:sipH} preserves gaussian states, with density matrices proportional to the exponential of a bilinear form in the creation-annihilation operators. Such density matrices can be parameterized by their matrices of two-point functions $G_{i,j}=\tr\big(\hat\rho_t \hat{a}^\dagger_i \hat{a}_j\big)$.
With simple algebra detailed in Appendix~\ref{app:2pt-dyn}, one obtains the noisy equation of motion for $G$ 
\begin{align}\label{eq:2pt-dyn}
& dG_{i,j} =  \big[-2G_{i,j}+\delta_{ij}(G_{i+1,j}+G_{i-1,j})\big]dt\\
&~~~~~~~~\; + i\big[ dW^i_t G_{i+1,j} +d\overline{W}^{i-1}_t G_{i-1,j} \nonumber\\
&~~~~~~~~\; ~~~~~~~~ -dW^{j-1}_tG_{i,j-1}-d\overline{W}^j_t G_{i,j+1}\big]\nonumber\\
&\; +\!\sum_{p\in\{0,N\}}\big[\alpha_p \delta_{i,p}\delta_{j,p}-\frac{\beta_p-\alpha_p}{2}(\delta_{i,p}+\delta_{j,p})G_{i,j}\big]dt,\nonumber
\end{align}
with the condition $\beta_p>\alpha_p$ needed to reach a steady-state regime at large times. We notice that Eq.~\eqref{eq:2pt-dyn} has the same form of that found for QSSEP in Ref.~\cite{Bernard2019}, up to a redefinition of the boundary terms as explained in Appendix \ref{app:steady-state}. This implies that the properties of the random matrix $G$ and of its fluctuations are analogous to those of the fermionic chain, despite the very different quantum nature of the particles. 

Denoting with $\textbf{[}\bullet\textbf{]}\equiv \lim_{t\to\infty}\E[\bullet]$ the steady-state statistical average over the stochastic realizations, one finds for instance the steady-state density profile along the chain
\be\label{eq:G1}
\bar{n}_i := \textbf{[}G_{i,i}\textbf{]}=\frac{n_a(N+b-i)+n_b(i+a)}{a+b+N}
\ee
and the second-order fluctuation of coherences ($i<j$)
\be\label{eq:G2}
\textbf{[}G_{i,j}G_{j,i}\textbf{]}^c=\frac{(n_b-n_a)^2(a+i)(N+b-j)}{(N+a+b)(a+b+N+1)(a+b+N-1)},
\ee
with definitions $a:=\frac{1}{\beta_0-\alpha_0}$, $b:=\frac{1}{\beta_N-\alpha_N}$, and $n_a:=\frac{\alpha_0}{\beta_0-\alpha_0}$, $n_b:=\frac{\alpha_N}{\beta_N-\alpha_N}$, see Appendix~\ref{app:2pt-dyn} and Ref.~\cite{Bernard2019} for details. In the thermodynamic limit $N\to\infty$, the boundary driving processes fix the local densities at the boundary sites to be respectively $n_a$ and $n_b$, without fluctuations.

Moreover, since \rev{Eq.~\eqref{eq:2pt-dyn}} is identical (up to irrelevant redefinition of the parameters) to those in QSSEP, the statistical properties of $G$ in QSSIP can be deduced from those in QSSEP. In particular, this implies that, with respect to the QSSIP invariant measure, the matrix $G$ is a representative of the ensembles of \textit{structured random matrices} introduced in \cite{Bernard2024}. Such classes of random matrices $M$ are characterized by the \hypertarget{eq:structured-RMT}{following properties} (called `axioms' in \cite{Bernard2024}):
\begin{enumerate}
\item[(i)] Local $U(1)$-invariance, meaning that, in law, $M_{i,j}=e^{-i\theta_i}M_{i,j} e^{i\theta_j}$ for any phases $\theta_i$ and $\theta_j$; 
\item[(ii)] Cumulants of cyclic products of matrix elements of order $n$, or ``loops'', scale as $N^{1-n}$, meaning that $\textbf{[}M_{i_1,i_2} M_{i_2,i_3}\dots M_{i_n,i_1}\textbf{]}^c={\cal O}(N^{1-n})$; moreover, these cumulants are continuous at coincident indices; 
\item[(iii)] To leading order in $N$, the expectation values of disjoint loops factorizes into the product of each loop expectation values; this still holds in the limit of touching loops.

\end{enumerate}
These three axioms have recently been completed by a fourth one \cite{ExtraBH2025} stating that
\begin{enumerate}
\item[(iv)] Cumulants of $r$ disjoint loops made altogether of $n$ entries of the matrix $M$ scale as ${\cal O}(N^{2-r-n})$; (the scaling in axiom (ii) is a peculiar case with $r=1$).
\end{enumerate}

\rev{As shown in Ref.~\cite{Hruza2023}, axioms (i–iii) follow from expected properties of the underlying noisy quantum dynamics at the mesoscopic scale. In particular, axiom (i) is a direct consequence of charge conservation of the model in Eq.~\eqref{eq:sipH}. Our choice of stochastic hopping amplitudes in Eq.~\eqref{eq:sipH}, modeled as complex Brownian motion increments, ensures that the system exhibits microscopic diffusion, see Appendix~\ref{app:2pt-dyn} and Ref.~\cite{Bernard2021}. Requiring the cumulants of local densities to be consistent with classical macroscopic fluctuation theory leads to the scaling property in axiom (ii), which, in turn, implies the factorization property stated in axiom (iii)~\cite{Hruza2023}. The fourth axiom is instrumental in establishing the self-averaging property of large deviation functions. Although a full mathematical proof is currently lacking, its validity has been tested in the QSSEP model using the exact solution found in Ref.~\cite{Bauer2024}. See Appendix~\ref{sec:self-avg} for details. There, we also present a complementary argument for self-averaging based on a stronger version of axiom (iii). Moreover, we remark that the validity of axioms \hyperlink{eq:structured-RMT}{(i–iv)} for loops of $G$ can be systematically checked from the quantum dynamics \eqref{eq:2pt-dyn} order by order, see Ref.~\cite{Bernard2019} for details.}\\

\rev{The axioms \hyperlink{eq:structured-RMT}{(i–iv)}} imply that fluctuations of the matrix elements satisfy a large deviation principle, with their size $N$ as large control parameter. In the large $N$ limit, they are fully parametrized by the expectation values of loops, with $x_k=i_k/N\in [0,1]$,
\be \label{eq:def-loopvalues}
g_n(x_1,\dots,x_n):=\lim_{N\to\infty} N^{n-1}\textbf{[}M_{i_1,i_2}M_{i_2,i_3}\dots M_{i_n,i_1}\textbf{]}^c
\ee


Since the matrix of two-point functions $G$, equipped with the QSSIP invariant measure, belongs to one of such ensembles, its statistical properties are fully encoded in its loop expectation values, see Eq.~\eqref{eq:def-loopvalues} with $M$ replaced by $G$. The latter coincide with those in QSSEP, which have been exactly computed by exploiting the connection of the problem to free probability~\cite{Bernard2021b,biane2022,Hruza2023}. For $n_a=0$ and $n_b=1$, they are recursively defined by
\be \label{eq:gn-free}
\sum_{\pi\in \text{NC}(n)} \prod_{p\in \pi} g_{|p|}(x_1,\dots,x_p)= \min(x_1,\dots, x_n)
\ee
where the sum is over the ensemble $\text{NC}(n)$ of non-crossing partitions of the ordered set $[1,\dots ,n]$.  For instance, for $n=1$ one has $\text{NC}(1)=\{1\}$ and thus $g_1(x)=x$; for $n=2$, $\text{NC}(2)=\{\{1,2\}\},\{\{1\},\{2\}\}$ yielding
\be
g_2(x_1,x_2)=\min(x_1,x_2)-x_1 x_2
\ee
(compare with Eqs.~\eqref{eq:G1}-\eqref{eq:G2} in the limit $N\to\infty$), and similarly to obtain higher orders. The expressions for $g_n$'s for other values of the boundary densities $n_{a/b}$ are obtained from this formula by simple shift and dilatation \cite{Hruza2023}.


\section{The mean classical process}

We now study the averaged process, which is actually a classical process. 
Despite the correspondence between the dynamics of the matrix of two-point functions in QSSEP and QSSIP, we shall show that the averaged classical process for the bosonic problem substantially differs from SSEP. 
This will \rev{allow} to test the conjecture that, to leading order, the large deviation functions of density fluctuations of such noisy many-body systems are given by the large deviation functions of the averaged classical process.  

Let $\bar\rho_t :=\textbf{[}\hat\rho_t\textbf{]}$ be the averaged density matrix. 
In mean, Eq.~\eqref{eq:lind} reduces to \cite{Bernard2018}
\be \label{eq:mean-flow}
\frac{d}{dt} \bar\rho_t={\cal L}_\text{bulk}(\bar\rho_t)+{\cal L}_\text{bdry}(\bar\rho_t),
\ee
with bulk mean Lindbladian defined as
\be
{\cal L}_\text{bulk}(\bullet)=\sum_{j=0}^{N-1} \Big(\hat{\ell}_j \bullet \hat\ell^\dagger_j+\hat{\ell}^\dagger_j \bullet \hat\ell_j-\frac12\{\hat\ell^\dagger_j\hat\ell_j+\hat\ell_j\hat\ell^\dagger_j,\bullet\}\Big),
\ee
and $\hat\ell_j\equiv\hat{a}^\dagger_j \hat{a}_{j+1}$. 
Writing $\bar{\rho}_t$ in the occupation number basis, this Lindblad evolution \eqref{eq:mean-flow} preserves diagonal density matrices, while off-diagonal elements vanish exponentially in time due to decoherence. Diagonal density matrices specify classical probability distributions and hence the evolution equation \eqref{eq:mean-flow} defines a classical process. 

Let us decompose the diagonal density matrices as $\bar\rho_t=\sum_{\vec{n}}P(\vec{n})\, \hat\Pi(\vec n)$ where the sum is over the classical particle configurations $\vec{n}=(n_0,n_2,\cdots,n_N)$ and $\hat\Pi(\vec{n})=\otimes_j |n_j\rangle\langle n_j|$ are the projectors on the occupation number eigenstates. The coefficients $P(\vec{n})$ define a probability measure on classical configuration, $0<P(\vec{n})<1$ and $\sum_{\vec{n}} P(\vec{n})=1$. Acting on pairs of adjacent projectors, $\hat\Pi_{j;j+1}^{n_j; n_{j+1}}:= \ket{n_j}\bra{n_j} \otimes \ket{n_{j+1}}\bra{n_{j+1}}$, the bulk mean Lindbladian generates the following Markov process, see Fig.~\ref{fig:SIP},
\begin{align}\label{eq:cl-bulk}
{\cal L}_\text{bulk}(\hat\Pi_{j;j+1}^{n; m} ) &= 
n(m+1)[\hat\Pi_{j;j+1}^{n-1;m+1} - \hat\Pi_{j;j+1}^{n;m}]  \nonumber\\
&+m(n+1)[\hat\Pi_{j;j+1}^{n+1;m-1}- \hat\Pi_{j;j+1}^{n;m}]
\end{align}
while the boundary terms ($p=0,N$) yields
\begin{align}\label{eq:cl-bdry}
&{\cal L}_\text{bdry}(\ket{n_p}\bra{n_p})=\alpha_p(n_p+1)\ket{n_p+1}\bra{n_p+1} \\
&\; +\beta_p n_p \ket{n_p-1}\bra{n_p-1}-\big[\alpha_p+(\alpha_p+\beta_p)n_p\big] \ket{n_p}\bra{n_p}.
\nonumber
\end{align}

This codes for transitions from one site to its neighborhood with probabilities depending on the occupations of the departure and arrival sites, reminiscent of the bosonic nature of the underlying quantum problem. Since the classical problem allows for the inclusion of multiple particles occupying the same site, we named it the \emph{Symmetric Simple Inclusion Process} (SSIP) following Refs.\cite{Giardina2010,Grosskinsky2011}. Recent studies have explored the asymmetric version of this classical transport model, see Refs.\cite{Minoguchi2023,Garbe2024}. \rev{In SSIP, the possibility of particles accumulating on a single site~--~a condensation-like phenomenon~--~is controlled by the injection and extraction rates $\alpha_p$, $\beta_p$ at the boundaries. For  $\beta_p > \alpha_p$, the system converges to a steady state with no condensation. The opposite regime, where boundary-induced condensation emerges, is briefly commented in Appendix~\ref{app:cond}.}\\

The relation between the quantum model QSSIP and its classical version SSIP can in particular be formulated via the moment generating function as
\be \label{eq:Zcl}
{\cal Z}_\mathrm{cl}[h]:= \langle e^{\sum_i h_i n_i}\rangle_\mathrm{ssip}=\tr\big(\bar\rho\, e^{\sum_i h_i \hat{n}_i}\big)
\ee
where $\hat{n}_i$ is the quantum number operator and $n_i$ the classical occupation number at site $i$, and $\bar \rho$ the averaged QSSIP density matrix. The classical cumulant generating functions is $\log {\cal Z}_\mathrm{cl}[h]$. Asymptotically, we expect that a large deviation principle holds so that
\be
{\cal Z}_\mathrm{cl}[h]\underset{N\to\infty}{\asymp} e^{N {\cal F}_\mathrm{cl}[h]}.
\ee

Notice that, a direct application of Wick's theorem reveals \cite{Bernard2019} the relation between the classical cumulants and the quantum loop expectation values, 
\be
\langle n_{i_1}\dots n_{i_k}\rangle_\mathrm{ssip}^c=N^{1-k} \sum_{\sigma\in S_k\setminus\mathbb{Z}_2} g_k(x_{\sigma(1)},\dots , x_{\sigma(k)})
\ee
for $i_1\neq \dots\neq i_k$ all distinct. The sum $\sigma\in S_k\setminus\mathbb{Z}_2$ runs over all permutations of $\{1,\dots ,k\}$ modulo cyclic permutations. This implies that the classical process is sensitive only to the sum of quantum loops over the $(k-1)!$ sectors connected by non-cyclic permutations,  and therefore it cannot alone determine the dynamics generated by Eq.~\eqref{eq:lind}.


\section{Large deviation function and its self-averaging behavior}

Our goal is now to determine the cumulant generating function of QSSIP, defined as the logarithm of
\be
{\cal Z}_\mathrm{qu}[h] := \tr\big(\hat\rho\, e^{\sum_i h_i \hat{n}_i}\big) ~,
\ee
with $h_i=h(i/N)$ with $h(x)$ a smooth test function.
Since the density matrix $\hat \rho$ is random, ${\cal Z}_\mathrm{qu}[h]$ is itself a random fluctuating variable. By construction its mean is the classical generating function, ${\cal Z}_\mathrm{cl}[h]= \textbf{[} {\cal Z}_\mathrm{qu}[h]\textbf{]}$.
Asymptotically at large $N$, we expect that a large deviation principle holds realization-wise, so that
\be \label{eq:Zqu}
{\cal Z}_\mathrm{qu}[h]\underset{N\to\infty}{\asymp} e^{N {\cal F}_\mathrm{qu}[h]} ~.
\ee
The cumulant generating function ${\cal F}_\mathrm{qu}[h]$ could a priori be realization dependent and fluctuating. But we shall argue that it is actually not random, at leading order in $1/N$, that is ${\cal F}_\mathrm{qu}$ is self-averaging, and thus it is equal to the classical cumulant generating function
\be \label{eq:Iqu=Icl}
{\cal F}_\mathrm{qu}[h] = {\cal F}_\mathrm{cl}[h] +\rev{ \mathcal{O}(1/\sqrt{N})} ~.
\ee
We shall furthermore give an exact expression for ${\cal F}_\mathrm{cl}[h]$.
Of course the sub-leading \rev{$\mathcal{O}(1/\sqrt{N})$} terms, which code for quantum correction, are random and fluctuating.\\

The argument goes as follows~--~further details are given in Appendix \ref{sec:self-avg}. Since realization-wise the QSSIP density matrix $\hat \rho$ is gaussian, proportional to the exponential of a quadratic form in the creation-annihilation operators, the generating function ${\cal Z}_\mathrm{qu}[h]$ can be expressed in terms of the matrix of two-point functions $G$. Namely, ${\cal Z}_\mathrm{qu}[h]$ can be written as a $N\times N$ determinant (see Appendix~\ref{app:bosonic-gauss-states} for a derivation)
\be\label{eq:MGF-det}
{\cal Z}_\mathrm{qu}[h] = \frac{1}{\det[ 1-Ge ]} = e^{-\tr\log[1-Ge]} ~,
\ee
with $e$ the diagonal matrix with entries $e_i=e^{h_i}-1$~--~i.e., $e= e^h-1$, and $h=\text{diag}(h_0,\dots,h_N)$. Let $M:=\log[1-Ge]$ so that, ${\cal Z}_\mathrm{qu}[h]=e^{-\tr M}$. It can be viewed as a formal power series of $Ge$. As explained above, $G$ is a realization of an ensemble of structured random matrices as defined by the axioms \hyperlink{eq:structured-RMT}{(i-iv)}, and so is $Ge$. It has been shown in  \cite{Bernard2024,ExtraBH2025} that this class of ensembles of random matrices is stable under non-linear transformations. Consequently, $M=\log[1-Ge]$ is also a realization of such structured random matrices. Axioms (iv) about the scaling of the cumulants of disjoints loops implies that the trace of any matrix in such ensemble is self-averaging. Thus $\tr [M] = \tr \log[1-Ge]$ is self-averaging, and
\be\label{eq:self-avg-qssip}
{\cal Z}_\mathrm{qu}[h] = e^{-\tr\log[1-Ge]} \underset{N\to\infty}{\asymp} e^{-\textbf{[} \tr\log[1-Ge]\, \textbf{]}}\ .
\ee
\rev{In other words, such self-averaging property implies that the random variable $\tr[M]$ becomes sharply peaked around its typical (or classical) value $\textbf{[} \tr[M]\textbf{]}$ in the limit $N \to \infty$. As a result, one has $e^{-\tr[M]} \asymp \textbf{[} e^{-\tr[M]} \textbf{]}\asymp e^{-\textbf{[} e\tr[M] \textbf{]} }$ with almost probability one.} Of course sub-leading terms in $N$ are not self-averaging. Consequently, we get that ${\cal Z}_\mathrm{qu}[h]$ is logarithmically asymptotically equal to its average, and 
\be
{\cal Z}_\mathrm{qu}[h] \underset{N\to\infty}{\asymp} e^{N {\cal F}_\mathrm{cl}[h]}
\ee
so that Eq.~\eqref{eq:Iqu=Icl} holds, ${\cal F}_\mathrm{qu}[h]={\cal F}_\mathrm{cl}[h]=:{\cal F}[h]$, with (we suppressed the indices `qu' and `cl' since they are equal) 
\be
{\cal F}[h] = -\lim_{N\to \infty} N^{-1} \textbf{[} \tr\log[1-Ge] \textbf{]} ~.
\ee
\indent 
A complete rigorous mathematical proof of this property would require a thorough and detailed analysis of the axioms \hyperlink{eq:structured-RMT}{(i-iv)} and their consequences. \\

This self-averaging property has been verified numerically, as shown in Fig. \ref{fig:num_check_self_avg}. The plot shows the variance of the quantity $ -\tr\log[1 - Ge]/N$ in the steady state of the QSSIP model with boundary densities $n_a = 1$ and $n_b = 0$, confirming the expected \rev{$\sim 1/\sqrt{N}$} scaling of the corrections. See Appendix~\ref{app:num} for details on the numerical implementation.\\

\begin{figure}[t]
    \centering
    \includegraphics[width=\columnwidth]{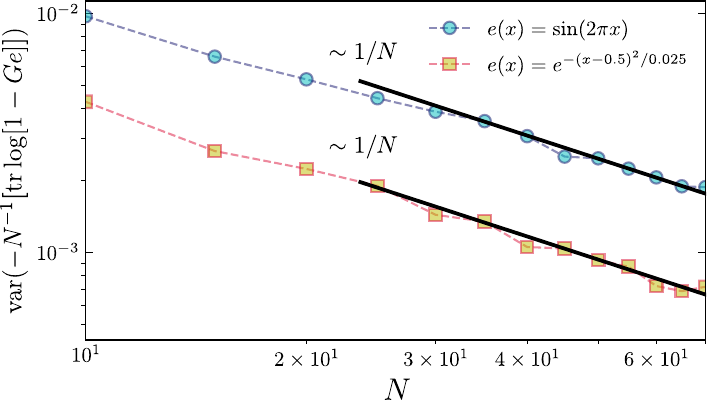}
    \caption{Numerical test of the self-averaging property \eqref{eq:self-avg-qssip}. Steady-state variance $\text{var}(O) := \textbf{[}(O - \textbf{[}O\textbf{]})^2\textbf{]}$ of the quantity $O[e]:=-N^{-1}\tr\log[1-Ge]$ in the QSSIP model, shown for two different arbitraty test functions $e(x)$. Steady-state convergence is verified from the one- and two-point functions (see Eqs.~\eqref{eq:G1}-\eqref{eq:G2}), with statistical averages taken over ensembles of at least $500$ samples. We set $\alpha_0=1$, $\beta_0=2$, and $\alpha_N=0$, $\beta_N=1$, corresponding to boundary site densities $n_a=1$ and $n_b=0$. Bold lines indicate the expected $\sim1/N$ scaling for reference.}
    \label{fig:num_check_self_avg}
\end{figure}

Furthermore, the analogy of ${\cal F}[e]=-\textbf{[}\tr\log(1-Ge)\textbf{]}/N$ with the generating functions analyzed in the context of spectrum of structured random matrices discussed in Refs.~\cite{Bernard2023,Bernard2024} demonstrates that ${\cal F}[e]$ is explicitly determined by the following extremization problem, with $e(x)=e^{h(x)}-1$, 
\be\label{eq:LDF-sip}
{\cal F}[h]=-\underset{a; b}{\text{extr}}\Big\{\!\!\int_0^1 \!\!\! dx\big[\log(1-e(x)b(x))+ b(x)a(x)\big] - {\cal F}_0[a]\Big\}
\ee
where ${\cal F}_0[a]$ is given in terms of the loop expectation values \eqref{eq:gn-free} by
\be\label{eq:F0}
{\cal F}_0[a]=\sum_{n\geq 1} \frac{1}{n} \int \prod_{k=1}^n\big(dx_k a(x_k)\big)\, g_n(x_1,\dots, x_n).
\ee
This exact formula resembles to a mean-field extension of the simplified case where sites are statistically independent. See Appendix~\ref{app:factorized-case} for details. 

The extremization problem in Eq.~\eqref{eq:LDF-sip} yields
\be\label{eq:extr}
b(x) =\frac{\delta {\cal F}_0[a]}{\delta a(x)}, \quad 
a(x) =\frac{e(x)}{1-b(x) e(x)},
\ee
and can be used to determine the density cumulants perturbatively in $e(x)$. The logic is that the order $e^{k-2}$ of $b(x)$ informs on the order $e^k$ of $a(x)$, and thus one can iterate the expansion of both functions self-consistently up to the desired order. 

As usual, the large $N$ asymptotic behavior (\ref{eq:Zcl},\ref{eq:Zqu}) of the cumulant generating function ensures that the probability of finding a given density profile $n(x)$ along the chain satisfies a large deviation principle,
\be
\text{Prob}[n(x)]\underset{N\to\infty}{\asymp} e^{-N {\cal I}[n]}.
\ee
with a large deviation function ${\cal I}[n]$, also called the rate function, equals to the Legendre transform of the generating function ${\cal F}[h]$,
\be\label{eq:rate-func-def}
{\cal I}[n]=\underset{h(\cdot)}{\text{extr}}\Big\{\int_0^1\!\! dx \, h(x)n(x) - {\cal F}[h]\Big\}.
\ee
Recall that $e(x)=e^{h(x)}-1$. The extremization condition yields $h(x)=\log\big(\frac{n(x)}{b(x)} \frac{1+b(x)}{1+n(x)}\big)$, from which one obtains the relations
\be\label{eq:extr2}
1-b(x)e(x)=\frac{1+b(x)}{1+n(x)}, \quad a(x)=\frac{n(x)}{b(x)}-\frac{1+n(x)}{1+b(x)}.
\ee
One can also verify that the extremization problem
\begin{align}\label{eq:rate-func}
{\cal I}[n]=\underset{a;b}{\text{extr}}&\Big\{\int_0^1 dx \Big[a(x)b(x) + n(x)\log\left(\frac{n(x)}{b(x)}\right) \\
& -[1+n(x)]\log\left(\frac{1+n(x)}{1+b(x)}\right)\Big]-{\cal F}_0[a]\Big\}
\nonumber
\end{align}
is equivalent to Eq.~\eqref{eq:rate-func-def}, as it reproduces the same conditions for the functions $a(x)$ and $b(x)$ as given in Eqs.~\eqref{eq:extr} or \eqref{eq:extr2}.
\\

Using techniques from free probability \cite{V97,Mi17,S19,Bi03}, we can present an alternative formula for the cumulant generating function $\mathcal{F}[h]$ or the large deviation function $\mathcal{I}[n]$ closer to the usual formula for SSEP, see Ref.~\cite{Derrida2007}. Namely, the generating function $\mathcal{F}[h]$ can be written as
\be\label{eq:LDF-sip-final}
{\cal F}[h]=\int_0^1 dx \ \log\frac{b'(x)}{1-e(x)b(x)}
\ee
with the function $b(x)$ solution of the following differential equation (recall that $e(x)=e^{h(x)}-1$)
\be\label{eq:diff-b}
b''(x)= e(x)\big[b(x) b''(x) + (b'(x))^2\big] 
\ee
with the boundary conditions $b(0)=0$ and $b(1)=1$. The proof of this formula is given in Appendix \ref{app:ldf-ssip}.
Consequently, the rate function ${\cal I}[n]$ can be written as
\begin{align}\label{eq:F-final}
{\cal I}[n]=\int_0^1 dx& \Big[-\log\left(b'(x)\right) + n(x)\log\left(\frac{n(x)}{b(x)}\right) \nonumber\\
& -[1+n(x)]\log\left(\frac{1+n(x)}{1+b(x)}\right)\Big].
\end{align}
By solving Eq.~\eqref{eq:diff-b} for a given test function $e(x)$, the exact cumulant generating function and large deviation function of density for the SSIP are then readily determined.\\

Expanding the generating function ${\cal F}[h]$ in power of the source field $h(x)$ yields the density cumulants. The non-local character of ${\cal F}[h]$ encodes for the long-range density correlations that occur in these cumulants.\\

A similar reasoning applies to QSSEP, with the expression of the QSSEP large deviation function in terms of $G$ differing from that in QSSIP by simple signs. Thus, the self-averaging property also applies to QSSEP. See Appendix \ref{sec:self-avg}. Sign differences in the large deviation functions of QSSIP (\ref{eq:LDF-sip},\ref{eq:F-final}) and QSSEP \cite{Bauer2024} can be systematically accounted for throughout the derivation.

\section{Entanglement}
Finally, we briefly comment on the calculation of entanglement in the quantum model~\eqref{eq:lind}. Let us introduce the $p$-th R\'enyi entropy of an interval $I\subset [0,1]$ as $S_I^{(p)}=\frac{1}{1-p}\log \tr (\hat\rho_I^p)$, with $\hat\rho_I$ the reduced density matrix $\hat\rho_I=\tr_{[0,1]\setminus I}(\hat\rho)$, for $p$ integer. For bosonic gaussian density matrices $\hat \rho$ as that of QSSIP, it can be written as (see Appendix~\ref{app:bosonic-gauss-states})
\be\label{eq:renyi-entropy}
S_I^{(p)}=\frac{1}{1-p} \tr \log\big[(1-G_I)^p-(G_I)^p\big]
\ee
with $G_I$ the truncation of $G$ to the interval $I$, i.e. $G_I=(G_{i,j})_{i,j\in I}$. Equivalently,
\be\label{eq:renyi}
S_I^{(p)}=\frac{\ell_I}{1-p}\int_0^1 d\sigma_I(\lambda)\log\big[(1-\lambda)^p-\lambda^p\big]
\ee
where $\ell_I$ is the length of $I$, and $d\sigma_I(\lambda)$ is the spectral distribution of the eigenvalues $\lambda \in [0,1]$ of $G_I$. We see that the R\'enyi entropy is different from that found in QSSEP \cite{Bernard2023}, due to the bosonic nature of the quantum particles. However, Eq.~\eqref{eq:renyi} is fully determined by the spectrum of $G_I$, which in turn is analogous to that of the fermionic problem. In Ref.~\cite{Bernard2023}, a discussion of such spectrum and the application to entanglement can be found. In particular, $S_I^{(p)}$ obeys a volume law as an echo of the long range correlations, typical of non-equilibrium systems.


\section{Summary and conclusions}

We considered a one-dimensional bosonic system with stochastic nearest-neighbor hoppings and boundary Lindblad drivings, as an analog to the QSSEP for fermionic particles. By exploiting the exact solvability of the quantum problem enabled by free probability, we derived analytical expressions for the steady-state cumulant generating function and the large deviation function of density profiles, see Eqs.~\eqref{eq:LDF-sip-final}-\eqref{eq:F-final}. This constitutes a step towards exploring a potential universal framework for describing fluctuations of transport or coherent phenomena  in quantum diffusive systems. 
 In performing our calculations, we identified the self-averaging property in Eq.~\eqref{eq:Iqu=Icl} as a key feature of such potential framework, providing elements of justification for the conjecture \cite{Hruza2023} on the classical nature of transport fluctuations in these systems. This self-averaging property in particular implies that the steady-state large deviation function of the quantum model coincides with that of the underlying classical system (sub-leading contribution are going to be different).
Future research directions include investigating the large deviation function of current fluctuations and verifying, or proving, that a similar self-averaging property holds. If this property persists, it would imply that classical transport properties, efficiently described by the macroscopic fluctuation theory~\cite{bertini2002macroscopic}, govern the behavior of fluctuations and rare events of transport in noisy diffusive quantum systems, up to sub-leading corrections. 
This conjectured classical nature of transport fluctuations does not apply to fluctuations of quantum coherent phenomena which are going to be dominated by their quantum character. Those would have to be described by a proper quantum extension of the macroscopic fluctuation theory \cite{Bernard2021}.

\medskip

\textit{Acknowledgements.}~---~We acknowledge L. Hruza and M. Albert for useful discussions and collaborations on closely related topics. This work was supported by the CNRS, the ENS, the ANR project ESQuisses under contract number ANR-20-CE47-0014-01. SS acknowledges support from the MSCA Grant No. 101103348 (GENESYS). This work has been partially funded by the European Union. Views and opinions expressed are however those of the author(s) only and do not necessarily reflect those of the European Union or the European Commission. Neither the European Union nor the European Commission can be held responsible for them.
\medskip 
\appendix

\section{Dynamics of the two-point function}\label{app:2pt-dyn}
In this appendix, we derive the dynamics of the two-point function $G_{i,j}=\tr\big(\hat\rho_t \hat{a}^\dagger_i\hat{a}_j\big)$ reported in Eq.~\eqref{eq:2pt-dyn}. We start from the equation of motion
\begin{align}
dG_{i,j}=&\langle i[d\hat{H}_t, \hat{a}^\dagger_i \hat{a}_j]\rangle-\frac12 \langle [d\hat{H}_t,[d\hat{H}_t, \hat{a}^\dagger_i \hat{a}_j]]\rangle 
\nn
&+ dt \langle {\cal L}_\text{bdry}^\star(\hat{a}^\dagger_i \hat{a}_j)\rangle
\end{align}
where $\langle\bullet\rangle\equiv\tr(\hat\rho_t\bullet)$ denotes the quantum expectation value, and ${\cal L}^\star_\text{bdry}$ is the dual boundary Lindbladian given by
\begin{align}
{\cal L}^\star_\text{bdry}(\bullet)=\sum_{p\in\{0,N\}}\Big[ &\alpha_p \Big(\hat{a}_p \bullet \hat{a}^\dagger_p-\frac12\{\hat{a}_p\hat{a}^\dagger_p,\bullet\}\Big)\nn
+&\beta_p \Big(\hat{a}^\dagger_p \bullet \hat{a}_p -\frac12\{\hat{a}^\dagger_p\hat{a}_p,\bullet\}\Big)\Big].
\end{align}
By evaluating the commutator,
\be
[\hat{a}^\dagger_m\hat{a}_n,\hat{a}^\dagger_i\hat{a}_j]=\delta_{n,i}\hat{a}^\dagger_m \hat{a}_j - \delta_{m,j} \hat{a}^\dagger_i\hat{a}_n
\ee
one finds for the unitary terms
\begin{align}\label{eqS:unitary1}
\langle i[d\hat{H}_t, \hat{a}^\dagger_i \hat{a}_j]\rangle=& i\big(d\overline{W}^{i-1}_t G_{i-1,j} + dW^i_t G_{i+1,j} \nn &- dW_t^{j-1} G_{i,j-1} - d\overline{W}^j_t G_{i,j+1}\big),
\end{align}
and
\begin{align}\label{eqS:unitary2}
\langle [d\hat{H}_t,[d\hat{H}_t, \hat{a}^\dagger_i \hat{a}_j]]\rangle=&
4dt G_{i,j} \nn &-2 dt \delta_{i,j}\big(G_{i+1,i+1}+G_{i-1,i-1}\big).
\end{align}
We see that the unitary part of the dynamics in Eqs.~\eqref{eqS:unitary1}-\eqref{eqS:unitary2} is analogous to QSSEP \cite{Bernard2019}. This is not surprising as the bulk dynamics of the chain is described by a single-particle problem, which is thus independent on the quantum statistics of the particles. The boundary Lindblad terms follow by evaluating a four-point function of bosonic field with Wick's theorem, reading as
\be\label{eqS:diss}
\langle{\cal L}^\star_\text{bdry}(\hat{a}^\dagger_i\hat{a}_j)\rangle=\sum_{p\in\{0,N\}} \big[ \alpha_p \delta_{i,p}\delta_{j,p} - \frac{\gamma_p}{2}(\delta_{i,p}+\delta_{j,p}) G_{i,j} \big].
\ee
with $\gamma_p\equiv\beta_p-\alpha_p$. We note that there is a sign difference w.r.t. the boundary terms found in QSSEP~\cite{Bernard2019}. Putting together Eqs.~\eqref{eqS:unitary1},\eqref{eqS:unitary2} and \eqref{eqS:diss}, we obtain the dynamics of the two-point function in Eq.~\eqref{eq:2pt-dyn}.


\subsection{Steady-state results} \label{app:steady-state}

To illustrate the analogy with the fermionic problem discussed in Ref.~\cite{Bernard2019}, we compute the steady-state mean density $\bar{n}_i=\textbf{[} G_{i,i}\textbf{]}$. Starting from Eq.~\eqref{eq:2pt-dyn}, one imposes the steady-state condition $\frac{d}{dt}\textbf{[}G_{i,i}\textbf{]}=0$, from which
\be\label{eq:fick}
\Delta_i \bar{n}_i +\sum_{p\in\{0,N\}} \delta_{i,p}\big(\alpha_p-\gamma_p\bar{n}_i\big)=0
\ee
with $\gamma_p>0$. Away from the boundary $i\neq 0,N$, the dynamics is purely diffusive and it fixes $\bar{n}_{i+1}=2\bar{n}_i-\bar{n}_{i-1}$, while condition at $i=0$ gives
\be
\bar{n}_1=\bar{n}_0\big(1+\gamma_0\big)-\alpha_0.
\ee
Thus, iterating this solution into the bulk, $\bar{n}_i= \bar{n}_0\big(1+i\gamma_0\big)-\alpha_0$. We can then fix the value of $\bar{n}_0$ using the boundary condition at $i=N$, obtaining
\be
\bar{n}_0=\frac{\alpha_0(1+\gamma_N N)+\alpha_N}{\gamma_0+\gamma_N+ N \gamma_0\gamma_N}.
\ee
Introducing,
\be\label{eqS:def-param}
a\equiv \frac{1}{\gamma_0}, \quad b\equiv \frac{1}{\gamma_N} \quad \text{and} \quad n_a\equiv\frac{\alpha_0}{\gamma_0},\quad  n_b\equiv \frac{\alpha_N}{\gamma_N},
\ee
one finds the result in Eq.~\eqref{eq:G1}. Note that, upon replacing $\gamma_p=\beta_p-\alpha_p$ with $\gamma^\text{qssep}_p=\beta_p+\alpha_p$ (and by changing the definitions in Eq.~\eqref{eqS:def-param} accordingly), the bosonic problem reduces to the fermionic one of Ref.~\cite{Bernard2019}. The same correspondence applies to the fluctuations of the coherences at all orders. 
\section{Self-averaging property of structured random matrices}\label{sec:self-avg}
In this appendix, we present additional arguments supporting the self-averaging property of structured random matrices $M$ that satisfy the axioms  \hyperlink{eq:structured-RMT}{(i-iv)} introduced in the main text. Specifically, we argue that the following property holds asymptotically
\be\label{eq:self-avg}
\textbf{[}e^{z\tr(M)}\textbf{]}\underset{N\to\infty}{\asymp} e^{z\textbf{[}\tr(M)\textbf{]}},
\ee
for any complex number $z$. We have $\textbf{[}\tr(M)\textbf{]}=\mathcal{O}(N)$.

As a simple check, one can show that the second-order cumulant $C_2:=\textbf{[}(\tr(M))^2\textbf{]}-(\textbf{[}\tr(M)\textbf{]})^2$ is sub-leading. We have $C_2=C_2^{(1)}+C_2^{(2)}$ with
\begin{align}
&C_2^{(1)}=\sum_{i}  \underbrace{\left(\textbf{[}M_{ii}^2\textbf{]}-\textbf{[}M_{ii}\textbf{]}^2\right)}_{{\cal O}(1/N)}\sim {\cal O}(1),\\
&C_2^{(2)}=\sum_{i\not= j} \underbrace{\left(\textbf{[}M_{ii}M_{jj}\textbf{]}-\textbf{[}M_{ii}\textbf{]}\textbf{[}M_{jj}\textbf{]}\right)}_{{\cal O}(1/N^2)}\sim {\cal O}(1),
\end{align}
as following from axiom (iv) with $r=2$ and $n=2$. Of course, this has been explicitly checked in QSSEP and QSSIP.

Similarly, axioms  \hyperlink{eq:structured-RMT}{(i-iv)} ensure that all higher order cumulants are subleading in $N$. We conjecture that these axioms are stable under polynomial non-linear transformations.
 
\subsection{Self-averaging in QSSEP}

We now (indirectly) prove that the self-averaging property holds for QSSEP. Using the QSSEP/SSEP correspondence, we know  \cite{Bernard2019,Bauer2024} that the moment generating function can be written as
\be\label{eq:ssep-MGF}
{\cal Z}^\text{ssep}[h]:=\textbf{[} \det[1+Ge]\textbf{]}\underset{N\to\infty}{\asymp} e^{N{\cal F}^\text{ssep}[e]},
\ee
with $e(x):=e^{h(x)}-1$ and $h(x)$ a smooth test function.

We shall show that the cumulant generating function in \eqref{eq:ssep-MGF} is indeed given by
\be\label{eq:self-avg-ssep}
{\cal F}^\text{ssep}[e]=N^{-1}\textbf{[}\tr(\log(1+Ge))\textbf{]}.
\ee

In Ref.~\cite{Bauer2024}, it has been proved using combinatorial arguments that ${\cal F}^\text{ssep}[e]$ appearing on the l.h.s. of \eqref{eq:self-avg-ssep} satisfies the extremization problem
\begin{align}\label{eq:ssep.tmp1}
{\cal F}^\text{ssep}[e]=\underset{a;b}{\text{extr}}\Big\{& \int_0^1\!\!  dx \left[\log(1+e(x)b(x))-a(x)b(x)\right] +\tilde{\cal F}_0[a]\Big\}
\end{align}
with
\be\label{eq:ssep.tmp2}
\tilde{\cal F}_0[a]=\sum_{n\geq 1} \frac{(-1)^{n+1}}{n} \int \prod_{k=1}^n\left(dx_k a(x_k)\right)\ g_n(x_1,\dots,x_n).
\ee
\indent

In parallel, in Ref.~\cite{Bernard2024} it has been proved that the function $F_z[\pm e]:=N^{-1}\textbf{[}\tr\log[z\mp Ge]\textbf{]}$~--~appearing on the r.h.s. of \eqref{eq:self-avg-ssep}~--~satisfies the extremization problem (setting $z=1$)
\begin{align}\label{bernard-hruza}
F_1[-e]=\underset{a;b}{\text{extr}} \Big\{&\int_0^1 \!\!dx\ \left[\log(1+e(x)b(x))-a(x)b(x)\right]\nonumber\\
&-{\cal F}_0[-a]\Big\}
\end{align}
with ${\cal F}_0$ given in Eq.~\eqref{eq:F0}. 

The last term can be rewritten as
\begin{align}\label{eq:ssep.tmp3}
{\cal F}_0[-a]&=\sum_{n\geq 1}\frac{(-1)^n}{nN} \sum_{i_1,\dots,i_n}  a_{i_1}\dots a_{i_n} \textbf{[}G_{i_1,i_2}\dots G_{i_n,i_1}\textbf{]}^c\nonumber\\[3pt]
&=\sum_{n\geq 1} \frac{(-1)^n}{n} \int_0^1 \prod_{k=1}^n\left(dx_k \ a(x_k)\right) \ g_n(x_1,\dots, x_n)\nonumber\\
&=-\tilde{\cal F}_{0}[a].
\end{align}
\indent
Comparing (\ref{eq:ssep.tmp1}-\ref{eq:ssep.tmp2}) and (\ref{bernard-hruza}-\ref{eq:ssep.tmp3}) proves the self-averaging property (\ref{eq:ssep-MGF}-\ref{eq:self-avg-ssep}) for QSSEP.

\subsection{Proof using a stronger version of axiom (iii)}
We now present a proof of the self-averaging property \eqref{eq:self-avg} based on the \emph{loop factorization hypothesis}, namely a stronger version of axiom (iii). This factorization hypothesis assumes that any products of closed
loops factorize up to subleading corrections in $1/N$:
\begin{align}
 & \textbf{[}G_{i_{1},i_{2}}G_{i_{2},i_{3}}\cdots G_{i_{n}}G_{j_{1},j_{2}}G_{j_{2},j_{3}}\cdots G_{j_{m}}\textbf{]}\label{eq:factorizationhyp}\nonumber\\
 & =\textbf{[}G_{i_{1},i_{2}}G_{i_{2},i_{3}}\cdots G_{i_{n}}\textbf{]}\textbf{[}G_{j_{1},j_{2}}G_{j_{2},j_{3}}\cdots G_{j_{m}}\textbf{]}\big(1+{\cal O}(1/N)\big).
\end{align}

It remains open to prove whether axioms  \hyperlink{eq:structured-RMT}{(i-iv)} also imply this stronger form of (iii). This property holds true up to third order fluctuations, as shown in Ref.~\cite{Bernard2019}.

Let $M:=-\log[1-Ge]$ be a structured random matrix satisfying \eqref{eq:factorizationhyp}. Our goal is to show \eqref{eq:self-avg},
where the $\asymp$ symbol means that the equality holds in the $N\to\infty$
limit for the log:
\begin{equation}\label{eq:self-avg2}
\lim_{N\to\infty}\log\textbf{[}e^{\tr(M)}\textbf{]}=\textbf{[}\tr(M)\textbf{]}.
\end{equation}By writing,
\be
\tr(M)=\sum_{n=1}^{\infty}\frac{1}{n}\sum_{j_{1},\cdots j_{n}}e_{j_{1}}\cdots e_{j_{n}}G_{j_{1}j_{2}}\cdots G_{j_{n}j_{1}},
\ee
and using the factorization hypothesis \eqref{eq:factorizationhyp} $k-1$ times, one obtains
\begin{equation}
\textbf{[}(\tr(M))^{k}\textbf{]}=\textbf{[}\tr(M)\textbf{]}^{k}\big(1+{\cal O}(1/N)\big)^{k-1}.
\end{equation}
This allows us to give an upper bound to $\textbf{[}e^{\tr(M)}\textbf{]}$:
\begin{align*}
&\textbf{[}e^{\tr(M)}\textbf{]}  =\sum_{k\geq 0}\frac{\textbf{[}(\tr(M))^{k}\textbf{]}}{k!} \nonumber\\
&\quad=\sum_{k\geq 0}\frac{\textbf{[}\tr(M)\textbf{]}^{k}}{k!}\big(1+{\cal O}(1/N)\big)^{k-1}\nonumber\\
&\quad \leq\sum_{k\geq 0}\frac{1}{k!}\left(\textbf{[}\tr(M)\textbf{]}\big[1+{\cal O}(1/N)\big]\right)^{k}\\
&\quad =\exp\left(\textbf{[}\tr(M)\textbf{]}\big[1+{\cal O}(1/N)\big]\right)
\end{align*}
so that 
\begin{equation}
\textbf{[}e^{\tr(M)}\textbf{]}\leq \exp\left(\textbf{[}\tr(M)\textbf{]}\big[1+{\cal O}(1/N)\big]\right).
\end{equation}
By convexity of the exponential, we also have the lower bound: 
\begin{equation}
e^{\textbf{[}\tr(M)\textbf{]}}\leq\textbf{[}e^{\tr(M)}\textbf{]}.
\end{equation}
Finally, taking the log and the limit $N\to\infty$ leads to the equality in Eq.~\eqref{eq:self-avg2}, thus completing the proof of \eqref{eq:self-avg}.
\rev{\section{Boundary-driven condensation}\label{app:cond}
\begin{figure}[t]
\centering
{\bf(a)} \hspace{2cm} {\bf(b)} \hspace{2cm} {\bf(c)}\\
\includegraphics[width=\columnwidth]{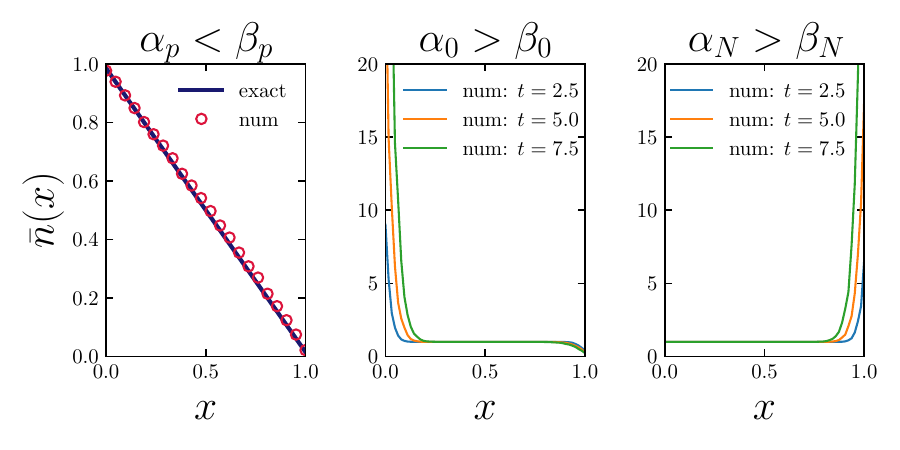}
\caption{\rev{Mean density profile as a function of position $x=i/N$ for different choices of boundary rates $\alpha_{0/N}, \beta_{0/N}$.
{\bf(a)}~--~Steady-state mean density $\bar{n}(x)$ for $\alpha_0 = 1$, $\beta_0 = 2$ (corresponding to $n_a = 1$) and $\alpha_N = 0$, $\beta_N = 1$ (i.e., $n_b = 0$). The exact result in \eqref{eq:G1} is shown with full line, while symbols denote the numerical data. {\bf(b)}~--~Numerical results for the mean density $\E[G_{i,i}]$ at different times for  $\alpha_0 = 2$, $\beta_0 = 1$ and $\alpha_N = 0$, $\beta_N = 1$ (i.e., $n_b = 0$). A boundary condensation emerges at the first site due to the injection-dominated left boundary.
{\bf(c)}~--~Same as panel (b) for $\alpha_0 = 1$, $\beta_0 = 2$ (i.e., $n_a = 1$) and $\alpha_N = 1$, $\beta_N = 0$. A boundary condensation develops at the last site due to the extraction-dominated right boundary. All numerical data are averaged over at least $100$ realizations.}}\label{fig:cond}
\end{figure}
We briefly comment on the boundary-driven accumulation (or condensation) of particles in the QSSIP model. To ensure the existence of a stationary state at large times, the injection and extraction rates at the boundaries must be balanced. Focusing on Eq.~\eqref{eq:fick}, we therefore obtain the stationariety condition at large $N$
\be
\alpha_p(1+\bar{n}_p) - \beta_p \bar{n}_p={\cal O}(1/N), \qquad p \in \{0, N\},
\ee
where $\bar{n}_p = \textbf{[}G_{p,p}\textbf{]}$ denotes the mean occupation at site $p$. Solving for $\bar{n}_p$, we find 
\be
\lim_{N\to \infty}\bar{n}_p = \frac{\alpha_p}{\beta_p - \alpha_p} \geq 0,
\ee
which implies that a steady state can only form if
\be
\alpha_p < \beta_p.
\ee
When this condition is satisfied, the system reaches a non-condensed steady state with finite densities throughout. Conversely, if $\alpha_p > \beta_p$, the occupation at site $p$ diverges and particles accumulate at the boundary, signaling the onset of boundary-induced condensation and the breakdown of stationarity. Numerical data obtained using the techniques discussed in Appendix~\ref{app:num} support this picture, as shown in Fig.~\ref{fig:cond}: for $\alpha_p > \beta_p$, boundary-driven condensation is clearly observed in the chain (see panels {\bf(b)}–{\bf(c)}).
}
\section{Bosonic Gaussian states}\label{app:bosonic-gauss-states}

In this appendix, we report useful manipulations of ${\cal Z}_\mathrm{qu}[h]$ obtained from Gaussian state techniques. Recall that for a family of bosons $\underline{a}=(\hat{a}_{i_1},\dots,\hat{a}_{i_N})$, the density matrix of a Gaussian state takes the form
 \be
 \hat\rho=\frac{1}{\cal Z} e^{\underline{a}^\dagger K \underline{a}},
 \ee
 with
 \be 
 {\cal Z}:=\tr\left(e^{\underline{a}^\dagger K \underline{a}}\right)=\frac{1}{\det[1-e^K]},
 \ee
 and with $K$ a $N\times N$ (hermitian) matrix. Since the map $K\mapsto \underline{a}^\dagger K \underline{a}$ defines a representation of $\mathfrak{gl}(N)$ in Fock space, the trace of the product of two group elements, $e^{\underline{a}^\dagger K \underline{a}} \ e^{\underline{a}^\dagger D \underline{a}}$, is given by $1/\det[1- e^K e^D]$.  As a consequence,
 \begin{align}\label{eq:boson-2pt}
 \tr\big(\hat\rho e^{\underline{a}^\dagger D \underline{a}}\big)&=  \frac{\det[1- e^K]}{\det[1-e^K e^D]}.
 \end{align}
 This can alternatively be written as
 \be \label{eq:last-bosonic-app}
  \tr\big(\hat\rho e^{\underline{a}^\dagger D \underline{a}}\big) = \frac{1}{\det[1-(e^D-1)G]} ,
 \ee
where we have identified the two-point function as
\be
 G:=\frac{e^K}{1-e^K},
\ee
since we can expand  $\tr\big(\hat\rho e^{\underline{a}^\dagger D \underline{a}}\big)$ as
\be
 \tr\big(\hat\rho e^{\underline{a}^\dagger D \underline{a}}\big)= 1+\tr(GD) + {\cal O}(D^2).
\ee

The expression \eqref{eq:MGF-det} for the moment generating function in the main text follows directly from \eqref{eq:last-bosonic-app}:
 \be
{\cal Z}_\mathrm{qu}[h]:=\tr\left(\hat\rho \ e^{\underline{a}^\dagger h \underline{a}}\right)=\frac{1}{\det[1-Ge]},
 \ee
where $e:=e^h-1$ and the test function $h=\text{diag}(h_1,\dots,h_N)$.\\

 With a similar algebra, the $p$-th R\'enyi entropy of an interval $I\subset[0,1]$ defined in Eq.~\eqref{eq:renyi-entropy}, can be written as
\begin{align}
S_I^{(p)}=&\frac{1}{1-p}\log\det\left[\frac{(1-e^{K_I})^p}{1-e^{p K_I}}\right]\nn
&=\frac{1}{1-p}\tr\log\left[(1-G_I)^p - (G_I)^p\right],
\end{align}
where $\hat\rho_I=\tr_{[0,1]\setminus I}(\hat\rho)=e^{\underline{a}^\dagger K_I \underline{a}}/{\cal Z}$ is the reduced density matrix in the interval $I$, and $G_I=(G_{i,j})_{i,j\in I}$ the associated two-point function.

\section{Numerical implementation}\label{app:num}
In this appendix, we provide further details on the numerical implementation of QSSIP, which was used to verify the self-averaging property in Fig.~\ref{fig:num_check_self_avg}. The stochastic evolution of the two-point function in Eq.~\eqref{eq:2pt-dyn} can be written as
\be\label{eq:discr-QSSIP}
G(t+dt)=e^{idh_t} G(t) e^{-idh_t} + dt\  {\cal L}_\text{bdry}[G(t)]
\ee
by expressing $d\hat{H}_t=\sum_{i,j=1}^N \hat{a}^\dagger_i \ dh_t \ \hat{a}_j$, leading to the $N\times N$ Hamiltonian matrix
\be
dh_t:=\begin{pmatrix}
0 & dW^1_t & \\
d\overline{W}^1_t& 0 & dW_t^2 \\
& d\overline{W}^2_t& 0 & dW_t^3\\
&& \ddots & \ddots & \ddots\\
&&&&0& dW_t^{N-1}\\
&&&& d\overline{W}^{N-1}_t & 0
\end{pmatrix}
\ee
and
\be
{\cal L}_\text{bdry}[G]_{i,j}\!\!=\sum_{p\in\{0,N\}}\!\!\!\left[\delta_{i,p}\delta_{i,j}\alpha_p-\frac{\delta_{i,p}+\delta_{j,p}}{2}(\beta_p-\alpha_p)G_{i,j}\right].
\ee
The discretization \eqref{eq:discr-QSSIP} leads to a straightforward numerical implementation of the QSSIP dynamics. The continuous stochastic process is approximated by selecting $dW_t^j$ at each time step as independent complex Gaussian variables with zero mean and variance $dt$ \cite{Bernard2021}. In our numerical calculations, we set the time step to $dt=10^{-2}$ and verified the stability of our results against different choices of $dt$.\\

Convergence to the steady state has been ensured from the one- and two-point functions (see Eqs.~\eqref{eq:G1}-\eqref{eq:G2}). For instance, the deviation from the one-point function is $\epsilon_i :=|\textbf{[}G_{ii}\textbf{]} -\bar{n}_i|$, and we checked that the error $\max\{\max_i (\epsilon_i),\max_i(\epsilon_i/\bar{n}_i)\} \leq 5\cdot 10^{-2}$.
\section{Generating function of the noninteracting problem}\label{app:factorized-case}
It is instructive to determine the cumulant generating function for the simplified problem where the site occupations $n_i$ are statistically independent. In this case, recall that the density matrix take the diagonal form $\hat\rho=e^{\sum_i \mu_i \hat{n}_i}/{\cal Z}$ with partition function ${\cal Z}=\prod_i \frac{1}{1-e^{\mu_i}}$. The moment generating function for such noninteracting problem is given by
\be
{\cal Z}_\text{free}[h]= \E\left[\prod_i \frac{1-e^{\mu_i}}{1-e^{\mu_i}e^{h_i}}\right]= \prod_i\E\left[ \frac{1-e^{\mu_i}}{1-e^{\mu_i}e^{h_i}}\right],
\ee 
where in the last equality we made use of the statistical independence of the sites to bring the average inside the product over sites. One can then relate the chemical potential $\mu_i$ to the quantum expectation value of the number operator $\hat{n}_i$ via
\be
n_i =\tr\big(\hat\rho \hat{n}_i\big)= \frac{e^{\mu_i}}{1-e^{\mu_i}},
\ee
and rewrite the moment generating function as
\be
{\cal Z}_\text{free}[h]=\prod_i\E\left[ \frac{1-e^{\mu_i}}{1-e^{\mu_i}e^{h_i}}\right]=\prod_i \frac{1}{1-e_i\bar{n}_i},
\ee
with $\bar{n}_i=\textbf{[}n_i\textbf{]}$, and $e_i\equiv e^{h_i}-1$. Introducing the cumulant generating function for the noninteracting problem as ${\cal Z}_\text{free}[h]\underset{N\to\infty}{\asymp} e^{N{\cal F}_\text{free}[h]}$, one finds 
\be\label{eqS:LDF-free}
{\cal F}_\text{free}[h]=N^{-1}\sum_i \log\frac{1}{1-e_i \bar{n}_i}=\int_0^1 dx \ \log\frac{1}{1-e(x)\bar{n}(x)}.
\ee
Comparing the free case in Eq.~\eqref{eqS:LDF-free} with Eq.~\eqref{eq:LDF-sip}, we observe that ${\cal F}[h]$ in the correlated case is given by a mean-field like formula. Indeed, the first term $\int dx \log\frac{1}{1-b(x)e(x)}$ resembles ${\cal F}_\text{free}[h]$ upon replacing the mean density $\bar{n}(x)$ with the effective density $b(x)$, which is determined from the non-coincident cumulants through the coupling to the external field $a(x)$ whose Boltzmann distribution is fixed by ${\cal F}_0$. A similar interpretation applies to the fermionic problem, as discussed in Ref.~\cite{Bauer2024}.

\section{Alternative formula for the SSIP large deviation function}\label{app:ldf-ssip}
In this appendix, we derive the alternative expression for the large deviation function of QSSIP given in Eq.~\eqref{eq:LDF-sip-final}, employing techniques from free probability theory. We introduce the R-transform (precisely, $1/\nu$ plus the R-transform)
\be
\mathscr{R}_{[a]}(\nu):=\sum_{\nu\geq 0}\nu^{n-1} g_n[a],
\ee
with $g_n[a]:=\int \prod_{k=1}^n(dx_k a(x_k)) g_n(x_1,\dots, x_n)$ and with the convention $g_0[a]\equiv 1$. $\mathscr{R}_{[a]}(\nu)$ generates the free cumulant of the classical variable $q_{[a]}(x):= \int_0^x dy\  a(y)$ with Lebesgue measure. Since the generating function ${\cal F}_0[\nu a]=\sum_{\nu\geq 1} \frac{\nu^n}{n} g_n[a]$ (see Eq.~\eqref{eq:F0}), it sastisfies the differential equation
\be\label{eq:tmp}
\nu\de_\nu{\cal F}_0[\nu a]=-1+\mathscr{R}_{[a]}(\nu).
\ee
Introducing also the associated moment generating function
\be
\mathscr{G}_{[a]}(\omega):=\sum_{n\geq 0} \omega^{-n-1} \phi_n[a],
\ee
one finds from free probability \cite{biane2022,Bauer2024}, 
\be
\phi_n[a]=\sum_{\pi\in \text{NC}(n)} \prod_{p\in \pi} g_{|p|}[a]=\int \prod_{k=1}^n\big(dx_k q_{[a]}(x_k)\big).
\ee
After simple algebra,
\be
\mathscr{G}_{[a]}(\omega)=\int_0^1 \frac{dx}{\omega - q_{[a]}(x)}.
\ee
Using the fundamental relation $\mathscr{R}_{[a]}\circ \mathscr{G}_{[a]}(\omega)=\omega$ \cite{V97,
Mi17, S19, Bi03}, Eq.~\eqref{eq:tmp} can be recast as
\be
-1+\nu\omega= \nu\de_\nu {\cal F}_0[\nu a]\ee
with the condition $\nu=\int_0^1 \frac{dx}{\omega -q_{[a]}(x)}$ relating $\nu$ and $\omega$. This equation can be readily integrated provided the boundary condition ${\cal F}_0[0]=0$. Restoring $\nu=1$, the final result reads as
\be\label{eq:I0-res}
{\cal F}_0[a]=\omega -1-\int_0^1 dx \log(\omega - q_{[a]}(x)),
\ee
with $\omega$ specified through $\int_0^1 \frac{dx}{\omega-q_{[a]}(x)}=1$. Eq.~\eqref{eq:I0-res} allows us to determine a differential equation for the function $b(x)$. Indeed, starting from its definition,
\be
b(x)\equiv \frac{\delta {\cal F}_0[a]}{\delta a(x)}=\int_0^x \frac{dx}{\omega- q_{[a]}(x)},
\ee
implying the boundary conditions $b(0)=0$ and $b(1)=1$. Next, by noticing that $1/b'(x)= \omega-q_{[a]}(x)$ and that $(1/b'(x))'=-a(x)\equiv-e(x)/(1-e(x)b(x))$, we obtain
\be\label{eq:diff-bis}
b''(x)-e(x)\big[b(x) b''(x) + (b'(x))^2\big]=0.
\ee
Furthermore, by rewriting the extremization condition \eqref{eq:extr} as $b'(x)[\omega-q_{[a]}(x)]=1$, one has that $\int_0^1 dx \ a(x)b(x)=1-\omega$. This identity allows us to recast Eq.~\eqref{eq:LDF-sip} in the form of \eqref{eq:LDF-sip-final}, i.e.,
\be 
{\cal F}[e]=\int_0^1 dx \ \log\frac{b'(x)}{1-e(x)b(x)}
\ee
with function $b(x)$ determined as solution of Eq.~\eqref{eq:diff-bis}.

\bibliography{bibliography}

\end{document}